# Probing Student Understanding With Alternative Questioning Strategies

Jeffrey M. Hawkins*[†], Brian W. Frank*[†], John R. Thompson*[†], Michael C. Wittmann*[†], Thomas M. Wemyss*[†]

*Department of Physics and Astronomy, University of Maine, Orono, ME 04469
[†]The Maine RiSE Center, University of Maine, Orono, ME 04469

**Abstract.** Common research tasks ask students to identify a correct answer and justify their answer choice. We propose expanding the array of research tasks to access different knowledge that students might have. By asking students to discuss answers they may not have chosen naturally, we can investigate students' abilities to explain something that is already established or to disprove an incorrect response. The results of these research tasks also provide us with information about how students' responses vary across the different tasks. We discuss three underused question types, their possible benefits, and some preliminary results from an electric circuits pretest utilizing these novel question types. We find that the answer students most commonly choose as correct is the same choice most commonly eliminated as incorrect. Also, given the correct answer, students can provide valuable reasoning to explain it, but they do not spontaneously identify it as the correct answer.

## INTRODUCTION

The results of physics education research are often used to inform science instruction. Research on students' misconceptions has led to curricular reform [1,2] and research on how students think about physics has led to new pedagogical approaches [3]. Because our assessment of students' ideas and thinking about physics has influences on curriculum and teaching, how we go about gathering data on student ideas and thinking is of fundamental importance.

Student ideas are often assessed by surveying students with written physics tasks. These surveys usually consist of asking a student for a correct answer to a question and possibly also an explanation of their reasoning [4, 5, 6].

We think methodologies for gathering information about student's ideas and thinking can be improved by varying the types of questions being used. As evidence for this claim, we present results from a trial study in which we asked three question types not commonly used in research or assessment tasks (novel question types) as well as a question asking for a correct response and explanation (*traditional*) as an electric circuits pretest.

## METHODOLOGY

For this trial of novel question types we distributed one question type to each student in a single large calculus-based introductory physics class at the University of Maine. Our four question types were all based on the same physical situation and each student answered only one question. The questions were distributed evenly in lecture with each question going to one quarter of the class.

The electric circuits pretest question we chose to modify was from *Tutorials in Introductory Physics* [6]. The traditional format, which asks for a correct response and an explanation, is shown in Figure 1 as the traditional version. We chose this pretest question because it was an appropriate content area that fit into our introductory physics course and because there was published research on students' ideas about this content [7]. This previous work guided the development of research tasks and the coding of students' responses, which were coded by a single researcher.

The pretest asks students to decide what happens to an indicator bulb (Bulb A) when a switch on a branch is closed. The correct response is that once the switch is closed, the total resistance of the circuit decreases, and therefore the total current flow though the battery increases; since Bulb A has the same current flow as the battery, Bulb A gets brighter.

When choosing which question types to use we looked at the results of previous research which showed that students do not have a good understanding of this content area [7]. Because of this, we chose to use novel question types that targeted the correct answer. All versions of the pretest we used are shown in Figure 1.

| Question | Responses |
|---|---|
| **Traditional Version** (N=22)<br>The circuit at right contains an ideal battery, three identical light bulbs, and a switch. Initially the switch is open.<br><br>After the switch closes:<br><br>Does the brightness of bulb A *increase, decrease,* or *remain the same?* Explain. 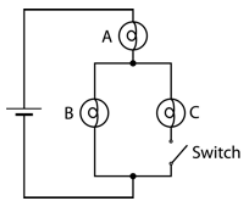 | 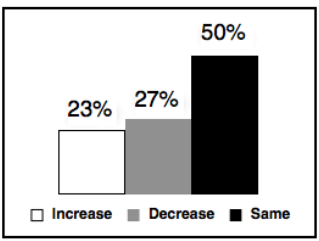 |
| **Incorrect Traditional Version** (N=16)<br>Imagine you are taking an exam with the question shown in the box below. You want to first eliminate one response you are pretty sure is incorrect. Which response would you eliminate? Why is that response the best one to eliminate?<br><br>The circuit at right contains an ideal battery, three identical light bulbs, and a switch. Initially the switch is open.<br>After the switch closes:<br>Does the brightness of bulb A *increase, decrease,* or *remain the same?* Explain. 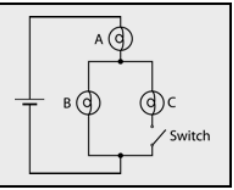 | 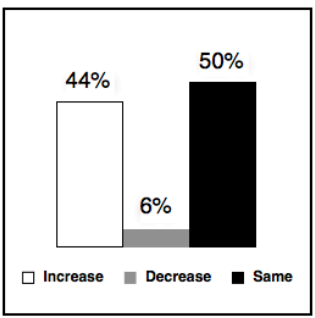 |
| **Given Increase Version** (N=26)<br>The circuit at right contains an ideal battery, three identical light bulbs, and a switch. Initially the switch is open.<br><br>After the switch closes, does the brightness of bulb A increase? Explain. 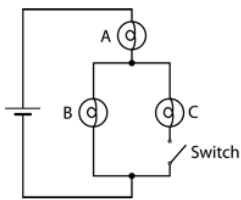 | 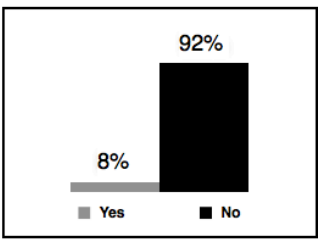 |
| **Given Correct Version** (N=26)<br>The circuit at right contains an ideal battery, three identical light bulbs, and a switch. Initially the switch is open.<br><br>After the switch closes, the brightness of bulb A increases. Explain. 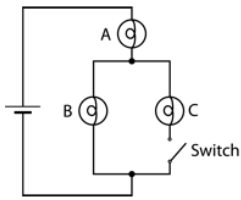 | 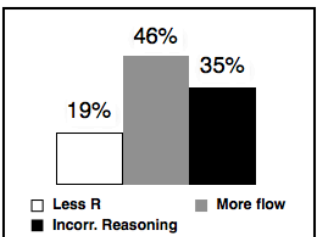 |

**FIGURE 1.** This figure shows the full text of each question type and the associated distribution of students' responses. The nine students who did not eliminate a response were removed from the analysis of *incorrect traditional*.

The first question type we chose to use was the *traditional* question type that asks students to select a correct response and explain their reasoning. Asking a traditionally formatted question provides a reference point to which our novel question types can be compared.

The second question type we chose to use was *incorrect traditional,* which asks students to select an incorrect response and explain their reasoning. Asking the *incorrect traditional* question type lets us know which answers students think are incorrect, and why they disbelieve them.

The third question type was *given increase*, which asks students if *increase* is a correct response and for them to explain their reasoning. Asking the *given increase* question type lets us know what students think of that answer even if they would not have chosen it as the correct one.

The fourth question type was *given correct*, in which we tell students that *increase* is the correct response and ask students to justify why it is the correct response. Asking the *given correct* question type lets us know if students have good reasoning for why the correct answer is correct even if they would have initially thought of it as incorrect.

# RESULTS

There are two ways that results from this set of question types can be analyzed. Results can be broken up by individual question type to see what the distribution of responses to each question is. Alternatively, results can be broken up by chosen response to see what the set of questions tells us about each possible response. We will start by briefly looking at the questions types individually, and then look at the possible responses individually.

## Individual Questions

The results of all four question types are displayed in Figure 1 and each question tells us something about what ideas students are bringing into the course. For the *traditional* question type, most students don't answer correctly, with *remains the same* as the most chosen response. The *incorrect traditional* question tells us that *increase* and *remains the same* are chosen as incorrect by the students, but that *decrease* is not often chosen as incorrect. The *given increase* question tells us that nearly all the students answering this question think *increase* is incorrect. The *given correct* question tells us that students can use formal reasoning about resistors, informal reasoning about flow, and some incorrect reasoning to justify or attempt to justify the correct response.

While these questions each provide us with valuable insights into students' ideas about electric circuits through their response distributions and provided reasoning, more interesting results are arrived at by looking across question types.

## Individual Responses

The *traditional* and *incorrect traditional* question types have all three responses as possible options and their results will be discussed with each possible response. The *given increase* and *given correct* question types focus on the response *increase* and will only be discussed in the section for that particular response.

### *Decrease*

The *decrease* option was given in the *traditional* and *incorrect traditional* question types. We see that 27% of one group chose it as a correct response, and 6% of another group chose it as an incorrect response. Combining these results indicates that the class as a whole views the response as possibly correct, but not definitively incorrect.

### *Remains The Same*

The *remains the same* option was also available on two questions. On the *traditional* question, *remains the same* was the response most commonly chosen as correct, with half the students choosing that response. On the *incorrect traditional* question, *remains the same* was the response most commonly eliminated as incorrect, with half the students responding this way. This interesting result has two possible interpretations. One suggests that half of the class that thinks *remains the same* is the correct response and the other half thinks it is an incorrect response. A different interpretation is that the different questions may elicit contrasting responses allowing for the possibility that one student may pick *remains the same* as correct or pick *remains the same* as incorrect depending on the question that student is asked. More research is needed to determine if either of these interpretations are correct.

The reasoning provided by students to justify eliminating or selecting *remains the same* also has an interesting contrast. The most common justification for why *remains the same* is a correct response is what McDermott referred to as "local reasoning" [7]. These students responses indicated that they where thinking about only a small region of the circuit. One student explained "The brightness of Bulb A will not change because it is not part of the system with the switch." The most common justification for why *remains the same* is an incorrect response was a holistic reasoning. These students reasoned that when you change one part of a circuit by closing a switch, the whole circuit changes. An example of this reasoning is: "I want to eliminate 'remain the same' because I feel that the brightness would have to change due to the change in circulation."

Combining these two question types allows us to see that *remains the same* has a distinctly different profile from *decrease*. While *decrease* seemed a little correct and even less incorrect to the class, *remains the same* seems both correct and incorrect to the students.

### *Increase*

On the traditional question, 23% of the students selected *increase* as the correct response. On the *incorrect traditional* question, 44% of students eliminated *increase*. These results show a difference between the responses *increase* and *decrease* as the percentage of the class that chooses them as correct is about the same, but *increase* is actively viewed as incorrect where *decrease* is not.

Looking at the results of the *given increase* question type we see only 8% of students said that the brightness of the bulb *increases*. There seems to be an inconsistency between 23% of students choosing *increase* as the correct response on the *traditional* question and only 8% saying the brightness *increases* on the *given increase* question type. One possible explanation is that some of the students selecting *increase* as the correct response simply thought it was the best option of the three even though they may not have thought of it as correct. These are different questions however, and the discrepancy may again be due to the influences of the different question types.

Lastly, when looking at the reasoning provided to justify why *increase* is the correct response on the *given correct* question, we find that most students have valuable reasoning about why *increase* is a correct answer. This is highly contrasted with the previous three questions, where students repeatedly indicated they do not view it as correct and, in fact, as incorrect. Of the responses to the *given increase* question, 19% of the students reasoned with formal reasoning that *increase* is correct because there is less total resistance when the switch is closed. An example of this type of reasoning is "The addition of Bulb C in a parallel circuit reduces the resistance in the system..." Another 46% of students responding to this question provided a more informal reasoning stating that once the switch was closed there was a second path for current to flow on and therefore the total current flow increased making the bulb brighter. An example of this type of response is "After the switch is closed, the charge will go through both the Left (B) side an the Right (C) side to increase the total charge Bulb A is receiving." The other 35% of the students just provided jargon or nonsense in their justification indicating that there are possibly some students who do not have any valuable reasoning for why *increase* is the correct response.

When combined, these question types give a nuanced view of what students think about *increase*. Instead of just knowing that some students think it is a correct response as the *traditional* question alone would tell us, we now have more detailed knowledge about how the class as a whole thinks about the response *increase* - some students choose it as correct, nearly half eliminate it as incorrect, very few say it is correct, and most can justify why it is correct.

## CONCLUSION

This set of questions act as a tool to give us a more in-depth understanding of students' ideas about the physical situation they are presented with. By using this set of questions, we have seen that *decrease* seems to be a reasonable answer to the students, as some say it is correct and very few say it is incorrect. We have found *remains the same* to be a bipolar answer, with half the students given the *traditional* question saying it is correct and half given the *incorrect traditional* question saying it is incorrect. Lastly we found that although only 8% of students indicate they think *increase* is correct when asked about it directly, 65% of students do have good reasoning to justify why it is correct when told it is correct.

This more detailed view of student ideas shows the value of using these novel question types to investigate students' ideas. This information has the potential to impact instruction in physics courses by providing more detail about the ideas students are bringing into physics courses.

These questions may also provide some insight into how students think about physics through contrasting responses to different questions. To further investigate what these questions tell us about students' thinking about physics we must further investigate how students interact with these questions. We plan to begin conducting this research through interviews and weekly pretests with varied question types.

## ACKNOWLEDGMENTS

We would like to thank all members of the Physics Education Research Laboratory for their valuable input regarding this research. We would also like to thank the Physics Education Research Topical Group and the Graduate Student Government at the University of Maine for providing partial funding for travel to PERC so this paper could be published. We would also like to thank Peter Shaffer and Mel Sabella for their insights.

## REFERENCES

1. P. S. Shaffer and L. C. McDermott, *Am. J. Phys.*, **73**(10), 921 (2005).
2. P.R.L. Heron, P.S. Shaffer, and L.C. McDermott, "Research as a Guide to Improving Student Learning: An Example from Introductory Physics," Invention and Impact, Proceedings of a Course, Curriculum, and Laboratory Improvement Conference, April 2004, Washington DC, (AAAS, 2005).
3. N. Podolefsky & N. Finkelstein, *Physical Rev. ST - PER*, **2**(2), 1-10 (2006).
4. R. K. Thornton & D. R. Sokoloff. *Am. J. Phys.*, **66**(4), 338-352 (1998).
5. L. Ding, R. Chabay, B. Sherwood, R. Beichner, *Physical Rev. ST – PER*, **2**(1), (2006).
6. L. C. McDermott & P. S. Shaffer, *Tutorials in Introductory Physics*, New Jersey: Prentice Hall, 2002.
7. L. C. McDermott & P. S. Shaffer, *Am. J. Phys.*, **60**(11), 994-1003 (1992).